\documentclass{aa}
\usepackage{graphicx,psfig}
\def\be{\begin{equation}}
\def\ee{\end{equation}}

\def\msun{\rm M_{\odot}}
\def\msunyr{\msun~{\rm yr^{-1}}}

\def\ergs{\rm \  \ erg \ s^{-1}}

\begin{document}

   \title{NGC~4258: a jet-dominated low-luminosity AGN?}


\author{Feng Yuan
        \and Sera Markoff\thanks{Humboldt Research Fellow} 
	\and Heino Falcke  \and Peter L. Biermann}

\offprints{F. Yuan; fyuan@mpifr-bonn.mpg.de}

\institute{Max-Planck-Institut f\"{u}r Radioastronomie, 
          Auf dem H\"{u}gel 69, D-53121 Bonn, Germany}

\date{Received 2001}

\abstract{ Low-luminosity AGNs (LLAGNs) are a very important class of
sources since they occupy a significant fraction of local
galaxies. Their spectra differ significantly from the canonical
luminous AGNs, most notably by the absence of the ``big blue bump''.
In the present paper, taking a typical LLAGN--NGC~4258--as an example,
we investigate the origin of their spectral emission.  The
observational data of NGC~4258 is extremely abundant, including water
maser emission, putting very strict constraints to its theoretical
models. The infrared (IR) spectrum is well described by a steep
power-law form $f_{\nu}\propto \nu^{-1.4}$, and may extend to the
optical/UV band. Up until now there is no model which can explain such
a steep spectrum, and we here propose a coupled jet plus accretion
disk model for NGC~4258.  The accretion disk is composed of an inner
ADAF (or radiatively inefficient accretion flow) and an outer standard
thin disk.  A shock occurs when the accretion flow is ejected out of
the ADAF to form the jet near the black hole, accelerating the
electrons into a power-law energy distribution.  The synchrotron and
self-Comptonized emission from these electrons greatly dominates over
the underlying accretion disk and can well explain the spectrum
ranging from IR to X-ray bands.  The further propagation of the
shocked gas in the jet can explain the flat radio spectrum of
NGC~4258.  Several predictions of our model are presented for testing
against future observations, and we briefly discuss the application of
the model to other LLAGNs. }

\maketitle

\section{Introduction}

The nearby active galactic nucleus (AGN) in NGC~4258 (M106) offers us
a rare opportunity to understand the powering of a class of AGNs
because of its precisely determined black hole mass and distance.
High resolution 22~GHz VLBI observations of the
water maser emission reveal a thin circumnuclear disk in Keplerian
orbit (Greenhill et al. 1995a; Miyoshi et al. 1995; Moran et al. 1995;
Herrnstein, Greenhill, \& Moran 1996).  The maser emission extends
from 0.13 pc to 0.26 pc, and the perfect Keplerian rotation curve
requires a central binding mass, $M$, of $(3.5 \pm 0.1) \times 10^7
\msun$ within 0.13 pc. This makes it one of the two strongest
supermassive black hole candidates to date (the other being our own
Galactic center, see Melia \& Falcke 2001 for a recent review).  By
measuring maser spot acceleration or proper motions (Greenhill et
al. 1995b; Herrnstein et al. 1999), the distance of the source can
also be directly determined independently of Hubble constant, and is
found to be $d=7.2 \pm 0.3~ {\rm Mpc}$.

In addition to the black hole mass and distance, the third critical
parameter for AGN modeling is the mass accretion rate $\dot{M}$.
The accretion rate in the maser disk can be determined 
from the observed water maser emission. The
projected positions of individual maser spots indicate that the maser
disk is significantly warped (Miyoshi et al. 1995; Greenhill et
al. 1995a). Such a warped disk will inevitably be illuminated
obliquely by the central X-ray source, and Neufeld, Maloney, \& Conger
(1994) show that X-ray irradiation of dense molecular gas is an
effective mechanism for generating powerful water maser emission. This
also naturally explains the association of water megamasers with AGNs
(Claussen \& Lo 1986; Braatz, Wilson, \& Henkel 1994).  By modeling
the molecular annulus as a standard thin accretion disk (SSD; Shakura
\& Sunyaev 1973) illuminated by the X-ray photons, Neufeld \& Maloney
(1995, hereafter NM95) infer the mass accretion rate through the disk
to be $\dot{M}_{\rm NM95}=7 \times 10^{-5} \alpha \msunyr$, here
$\alpha \la 1$ is the conventional viscosity parameter.

However the NM95 accretion rate is obtained from modeling the masering
disk, which is very far away from the hole since 0.13~pc $\simeq 40000
r_s$ with $r_s\equiv 2r_g=2GM/c^2$ is the Schwarzschild radius of the
black hole.  The viscous timescale at 0.13~pc is extremely long, $\sim
10^9\alpha_{-1}^{-1}(200{\rm K}/T)(r/0.2~{\rm pc})^{1/2}$~yr,
therefore, it is very unlikely that accretion has been steady during
such period (Gammie, Narayan, \& Blandford 1999, hereafter GNB99).  It
is thus likely that the accretion rate at the inner region of the
accretion disk can has a different value from NM95 (GNB99).

There is a plethora of observational data for NGC~4258, ranging from
the radio, IR, optical/UV, to X-ray bands, as we will review in
Sect. 2. These broadband data indicate that NGC~4258 is very
under-luminous, $L/L_{E}\approx 10^{-4}$, where $L_{\rm E}$ is the
Eddington luminosity, making NGC~4258 an ideal candidate for an
advection-dominated accretion flow (ADAF; see reviews by Narayan,
Mahadevan, \& Quataert 1999 and Kato, Fukue \& Mineshige 1998).  In
particular, since a thin disk is observed at the region out of
$\sim0.13~{\rm pc}$, a truncated ADAF model is viable in which an ADAF
connects at a transition radius $r_{\rm tr}$ to an outer SSD. This
would naturally explain why there is no maser emission inside 0.13~pc.
Several transition mechanisms have been proposed (e.g., Meyer
\& Meyer-Hofmeister 1994; Honma 1996; Czerny, Rozanska, \& Zycki 2000;
Spruit \& Deufel 2002), but a reliable prediction for the value of
$r_{\rm tr}$ is still lacking.

Lasota et al. (1996) assume that such a hybrid accretion disk is
responsible for the spectrum of NGC~4258. This model is later
refined in GNB99.  The X-ray spectrum of NGC~4258, which is hard to
explain by a SSD alone, can be well explained by the ADAF. The IR
spectrum of NGC4258 was not certain at the time when GNB99 was
published, although GNB99 could well explain the IR flux which was
available at that time. The IR spectrum in this model is produced by the
outer SSD, therefore a very hard IR spectrum with the classical
form of $f_{\nu}\propto \nu^{1/3}$ is predicted.  The required accretion
rate $\dot{M} \sim 0.01 \msunyr$ (GNB99).  This is more than two
orders of magnitude higher than the NM95 rate.

After the publication of GNB99, new IR observations determined a
better spectrum (see next section for details), which is in serious
conflict with the prediction of GNB99. The detected spectrum is very
steep, with spectral index $s = 1.4 \pm 0.1$ ($f_{\nu}\propto
\nu^{-s})$. On the other hand, when fitting the spectrum of NGC4258
(or any source), an ADAF model usually does not take into account the
role of the jet, which is believed to occur in the innermost region of
the accretion disk. In some cases, the emission from the jet may
dominate over the underlying accretion disk in sources such as
blazars, M87 (Wilson \& Yang 2002), and some classes of X-ray binaries
(Markoff, Falcke, \& Fender 2001).

In the present paper, we propose an alternative model to explain the
spectrum of NGC~4258. We suggest that the main emission of NGC~4258
comes from the base of its jet rather than the accretion disk. In next
section, we will present a review of the relevant observations of
NGC~4258.  Sect. 3 presents our model and explanations for the
spectrum of NGC~4258, along with some theoretical predictions. The
application of our model to LLAGNs in general is discussed in the last
section.

\section{Review of Observations}

\subsection{Radio}

VLA observations reveal a compact radio source within $0.^{"}1$ of the
maser emission, with a $\sim 3$~mJy flux density at 2~cm, and a
spectral index $s$ of $\sim -0.4$ (Turner \& Ho 1994), indicated by
the short ``slanted'' line in Fig.~1.
 
Herrnstein et al. (1998) report that the 22~GHz radio emission traces
a jet-like structure oriented along the rotation axis of the maser
disk, and composed of distinct northern and southern components with
flux density of $2.5-3.5$~mJy and 0.5~mJy, respectively.  Because the
relative position of the center of mass of the molecular disk can be
measured to a fraction of a milliarcsecond with VLBI, NGC~4258
provides a rare opportunity to circumvent the ambiguity associated
with core-jet emission.  The centroids of the northern and southern
components appear at $0.35-0.46$~mas and $-1.0$ mas from the disk
center, respectively.  Both the flux and centroid of the northern
component vary significantly with time. More interestingly, Herrnstein
et al. (1998) report a 3~$\sigma$ upper limit of 220~$\mu$Jy on any
22~GHz continuum emission coincident with the central mass of the
NGC~4258 maser disk.  This result places an extremely strict
constraint on theoretical models of NGC~4258, and is the main
motivation for GNB99's revisiting of the ADAF model after Lasota et
al.'s (1996) work. GNB99 argue that such an upper limit requires the
SSD-ADAF transition radius $r_{\rm tr} \sim (10-100)r_s$.

\subsection{Optical/UV}

Wilkes et al. (1995, hereafter W95) have measured the 5500~\AA ~flux
toward NGC~4258 in polarized light. This emission presumably arises in
the central engine and has been scattered into our line of sight.
They find that the spectral index is well-fit by a power law with
$s=1.1 \pm 0.2$.  This value agrees well with the argument of St\"uwe,
Schulz, \& H\"uhnermann (1992) that a continuum with $s \sim 1.2$ can
produce sufficient ionizing power to explain the observed strength of
the optical emission lines in NGC~4258.  The corresponding 5500~\AA~
central engine luminosity is highly sensitive to the type of
scattering screen invoked.  W95 proposed two candidates for the
polarization: scattering by electrons in an H~II region surrounding
the nucleus, and scattering by dust grains located above and below the
molecular disk. The estimated luminosities are $L_{5500} \sim 10^{39}
{\rm erg s^{-1} \AA^{-1}}$ and $10^{37} {\rm erg s^{-1} \AA^{-1}}$,
respectively.  We denote both of them in Fig.~1 with empty circles
because they are not very precise.  However, we think that the latter
is more likely because connecting the near-IR data to the former point
would result in an unbelievably hard spectrum, in serious conflict
with the observed polarization spectrum.

\subsection{Infrared} 

Chary \& Becklin (1997) present J, H, and K band Keck observations.
Using the J band surface brightness profile as a baseline, they obtain
the J-band subtracted H and K bands image which is consistent with a
unresolved point source. The nucleus is then found to be very red.
Such redness can be attributed to either a red central source or dust
extinction of starlight. They argue for the former by showing that
this color cannot be explained by uniform foreground dust extinction,
assuming a late-type stellar population typical of the bulge. Assuming
that the intrinsic spectrum has a similar power-law form with the
optical/UV polarization spectrum obtained by W95, which gives $f_{\nu}
\propto \nu^{-1.1 \pm 0.2}$, they deduce the extinction and
luminosity.

Obviously such a spectral result, especially the value of the
power-law spectral index, is not so certain.  For example, a power-law
fit of $f_{\nu}\propto\nu^{1/3}$ also works well, as pointed out by
GNB99.  Using HST (NICMOS), the 5.1 m Hale telescope and the Keck II
telescope (MIRLIN), Chary et al. (2000) went on to present
high-resolution images from 1 to 18 $\mu$m, placing a much better
constraint on the spectrum.  They find that the mid-IR data before
extinction-correction share a similar spectrum with the optical/UV
polarization results. Since mid-IR suffers relatively little dust
extinction, such a similarity strongly suggests that the IR spectrum
has approximately the same power-law form as the optical/UV
polarization spectrum.  To make sure, they tested four predictions for
the 3.5 $\mu$m flux based on four different power-law spectral
indexes, including $s=-1/3$.  They find that only a steep power-law
with a similar spectral index as the polarization spectrum of W95 can
give a correct prediction with their observation (R. Chary, private
communication). In addition, the extinction coefficient obtained from
such a steep spectral fit across all the IR wavebands is consistent
with the extinction derived from the N-band data point after including
the 9.7 micron silicate feature (R. Chary, private communication).
They therefore conclude that the intrinsic extinction-corrected IR
spectrum of NGC~4258 can be well fitted with a steep power-law form,
and their best fit extinction-corrected spectrum goes as $f_{\nu}
\propto \nu^{-1.4 \pm 0.1}$. Considering W95, such a spectrum is very
likely to extend to the optical/UV band.  The data is listed in Table
1 and shown by filled circles in Fig. 1.

\begin{table*}
\caption[]{Radio, infrared, and optical observational data for NGC~4258.}
\begin{tabular}{lllrrrccr}
\hline\noalign{\smallskip}
\multicolumn{1}{l}{Frequency(Hz)}  & \multicolumn{1}{c}{Flux(mJy)} &
\multicolumn{1}{c}{Reference} \\
\noalign{\smallskip}\hline\noalign{\smallskip}
$5-15\times 10^9$      & 3(s=-0.4$^a$) & 1\\
$2.2 \times 10^{10}$   & 2.5-3.5, $<0.22^b$& 2\\
$1.68\times 10^{13}$   & 435 & 3 \\
$2.40\times 10^{13}$   & 270 & 3 \\
$2.86\times 10^{13}$   & 195 & 4 \\
$8.70 \times 10^{13}$  & 46  & 3\\
$1.36 \times 10^{14}$  & 25  & 3\\
$1.88 \times 10^{14}$  & 16  & 3\\
$2.73 \times 10^{14}$  & 10  & 3\\
$5.45 \times 10^{14}$  & 1.7,170$^c$& 5\\
\noalign{\smallskip}
\hline
\noalign{\smallskip}
\end{tabular}
\label{spectra}

Notes and references:\\
$^{ a}$: s=-0.4 is the spectral index ($f_{\nu}\propto \nu^{-s}$)\\
$^{ b}$: 2.5-3.5mJy is the flux detected from the jet, while 0.22mJy is
the upper limit detected from the location coincident with the central mass
of the maser disk\\
$^{c}$: the two values correspond to two scattering models\\
(1) Turner \& Ho 1994;
(2) Herrnstein et al. 1998;
(3) Chary et al. 2000;
(4) Rieke \& Lebofsky 1978;
(5) W95
\end{table*}

Such IR and optical/UV spectral results, combining with the radio
data, give an estimate the bolometric luminosity of NGC~4258 of
$L_{bol} \approx 8 \times 10^{41}- 3 \times 10^{42}\ergs$. Spinoglio
\& Malkan (1989) suggest that the 12 $\mu m$ luminosity is typically
about 1/5 of the bolometric luminosity of AGNs, independent of whether
the emission is thermal or nonthermal in origin. This argument gives
$L_{bol}\sim 10^{42}\ergs$ for the mid-IR data of Chary et al
(2000). The consistency between the two estimates again suggests that
the steep power-law spectrum extends to the optical/UV, and that there
is no big-blue-bump as usually found in luminous AGNs.  The steep
power-law spectrum, $f_{\nu} \propto \nu^{-1.4}$, found throughout the
IR and possibly extending to optical/UV strongly suggests a
non-thermal origin, and that the dust in NGC~4258 is likely to be cool
and contribute only at wavelengths longer than mid-IR (Chary et
al. 2000).
 
Such a steep IR spectrum is very unusual compared to the canonical
luminous AGNs but seems to be common in LLAGNs, as we will illustrate
in Sect. 4. It provides a strict constraint to the theoretical model
of NGC~4258\footnote{Chary et al. (2000) did try to fit such a
spectrum using an ADAF model, but their calculation is too
simplified. A more exact ADAF calculation predicts an exponentially
decreasing spectrum at IR band, as shown by, e.g., the dot-dashed line
in Fig. 1 in the present paper.}.  In particular, this is in serious
conflict with the prediction of $f_{\nu}\propto\nu^{1/3}$ by GNB99, as
discussed in Sect. 1.

\subsection{X-rays}

Several telescopes have observed the X-ray emission of NGC~4258 in
recent years, from ASCA nearly eight years ago to XMM-{\em
Newton} reported this year. The reader can refer to Table 1 in
Pietsch \& Read (2002) for the values of spectral parameters. An
interesting result we find is that the observed X-ray spectral indexes
differ greatly for different observations, ranging from $s=0.64$
(Pietsch \& Read 2002) to $s= 1.11$ (Fiore et al. 2001). We put these
two most extreme results in Fig. 1.  While one reason for the
discrepancy of the spectral index is likely due to the instruments, it
is possible that the spectrum is intrinsically variable with time.

Reynolds, Nowak, \& Maloney (2000) report their ASCA detection of the
iron K$\alpha$ line.  They find the line is narrow\footnote{The
detected iron K$\alpha$ line is moderately strong. This is somewhat
unusual since among LLAGNs the iron line is either extremely weak or
absent (Terashima et al. 2002).  The iron line in NGC~4258 may come
from, e.g., the warping molecular disk.}, which constrains the inner
radius of the SSD to be  $>100r_g$.  Note in GNB99 the SSD-ADAF
transition radius is $r_{tr} \sim (20-200)r_g$, so it only marginally
satisfies the iron line observation.

In the {\em Beppo}SAX observation of Fiore et al. (2001), large
amplitude (100\%) variability is observed on timescales of a few tens
of thousands of seconds, while variability of $\sim 20 \%$ is observed
on timescales as short as 1 hr.  The size of the X-ray source is then
constrained to be $< 100 r_g$.

\subsection{Summary of Observations}

In short, the spectrum of NGC~4258 is quite different from the usual
luminous AGNs.  As discussed above, the data puts very strict
constraints on the theoretical models and challenges our understanding
of the powering mechanism in NGC~4258. On the other hand, we note that
NGC~4258 is somewhat analogous to Sgr A$^*$: both sources have flat
radio and steep IR spectra, and both are low luminosity sources with
the luminosity peaked at sub-mm or IR rather than optical as in
luminous AGNs. However, the bolometric luminosity of NGC~4258 is about
$10^4$ times higher than Sgr A$^*$ in units of Eddington luminosity,
and the jet in NGC~4258 is much stronger than in Sgr A$^*$, which if
present has not yet been detected.  The spectral similarity between
the two sources does, however, suggest a common physical mechanism. A
jet-ADAF model has been found to be very successful at explaining the
spectrum of Sgr A$^*$ from radio to hard X-ray (see Yuan, Markoff, \&
Falcke 2002 for details; hereafter YMF02)\footnote{ Recently the
presence of a four-month cycle in the radio variability of Sgr A$^*$
was reported (Zhao, Bower, \& Goss 2001; see also Yuan \& Zhao
2002). By analogy with the plateau state of GRS 1915+105, it is
suggested that such quasi-periodic variability is very likely to be
produced through a jet (Zhao, Bower, \& Goss 2001). If verified, this
observation would strengthen arguments for a jet component in models
of Sgr A$^*$.}. We consider that the same type of model may be
applicable to the spectrum of NGC~4258 as well. In this model, most of
the emission comes from the base of the jet rather than the accretion
disk.  The much higher luminosity of NGC~4258 compared to Sgr A$^*$ is
then due to its more powerful jet.
 
\section{Jet Model for NGC~4258}

The physical scenario of our model is as follows. The accretion flow
outside of 0.13 pc (or smaller) is described as a SSD. At a certain radius, a
SSD-ADAF (or more generally a radiatively inefficient accretion flow)
transition occurs.  Since the accretion onto the black hole must be
transonic, the accretion flow becomes radially supersonic after
passing through a sonic point $r_{\rm sonic} \la 10 r_g$ (e.g.,
Narayan, Kato, \& Honma 1997).  When the radially supersonic accretion
flow is subsequently ejected out of the disk to form the jet, a
standing shock occurs due to the bending, as suggested by the general
relativistic MHD numerical simulations of jet formation (Koide,
Shibata \& Kudoh 1998; Koide et al. 2000). After the shock, the
kinetic energy of the pre-shock accretion flow will be transferred
into the internal energy of the post-shock plasma and the temperature
of the flow will be increased nearly to the thermal temperature $\sim
10^{12}/(r/r_g){\rm K}$, since the energy loss from the pre-shock ADAF
is very small and can be neglected.  Most of the emission from such a
disk-jet system then stems from the high-temperature post-shock
plasma. We find that the shock is radiative, i.e., the radiative
cooling timescale is much shorter than the characteristic shock
propagation time. We will show that the synchrotron emission will be
responsible for the steep IR spectrum of NGC~4258 while its inverse
Compton emission can explain the X-ray spectrum.
The flat radio spectrum is explained by the emission from the outer 
part of the jet.

\subsection{The Shock at the Jet Base}

We denote the shock (or jet) radius as $r_0$. Obviously, $r_0$ should
be less than the sonic radius of ADAF, $r_0 \la r_{\rm sonic}$.  Some
accretion flow outside of $r_{\rm sonic}$ may also go into the jet,
but since it will not be shocked, its radiative contribution is small
enough to be neglected.

We calculate the temperature of the post-shock plasma from the
parameters of the pre-shock flow using the shock transition
conditions.  The post-shock equivalent temperature is determined by
\be T^{eq}_2 \simeq \frac{(\Gamma+1)}{2(\Gamma+1)^2}
\frac{m_p}{k}v_1^2(r_0)=\frac{1}{5} \frac{m_p}{k}v_1^2(r_0).  \ee The
total thermal energy density of the post-shock plasma is then \be e
={3}nkT^{eq}_2.  \ee Here the strong shock approximation and a
parallel shock configuration are assumed. The adiabatic index
$\Gamma=1.5$ since the protons are nonrelativistic while the electrons
are relativistic, $n(=n_p=n_e)$ is the particle number density,
$v_1(r_0)$ is the pre-shock velocity of the accretion flow at the
shock location $r_0$. We obtain $v_1(r_0)$ by solving
the global solution for ADAFs.

However, since we do not know what fraction of the accretion flow will
go into the jet, and because hydrostatic equilibrium in the vertical
direction of the disk does not hold at the jet formation region, it is
hard to determine the density of the post-shock plasma. As in YMF02,
here we adopt the mass loss rate in the jet, $\dot{M}_{\rm jet}$, as a
free parameter. Then the density of the post-shock plasma follows
from \be \dot{M}_{\rm jet}=4\pi r_0^2 n_p m_p v_2.  \ee Here $v_2$ is
the velocity of the post-shock flow and we calculate its value from
the parameters of the pre-shock accretion flow using the shock
transition conditions. Obviously $\dot{M}_{\rm jet}$ should be smaller
than the disk accretion rate $\dot{M}$.

We assume that electrons are accelerated in the shock to a power-law
energy distribution, with a minimum Lorentz factor $\gamma_{\rm min}$:
\be N(\gamma_e) = N_0 \gamma_e^{-p} \hspace{1cm} (\gamma_e \geq
\gamma_{\rm min}).  \ee Generally $p\approx 2.5$ is assumed in the
literature for the {\em injected distribution} in the case of strong
relativistic shock acceleration, e.g., Gamma-ray Bursts (GRBs).  But
in the present case, the shock is sub-relativistic and only moderately
strong. For such a shock, we still know little about the form of the
energy distribution of accelerated/heated electrons (e.g., Draine \&
McKee 1993).  We conjecture that $p$ is likely somewhat larger, or
even a thermal distribution is possible in the extreme case.  We set
$p=2.8$ in the present paper to relate it to the observed IR spectral
index $s=1.4$
\footnote{In the case of Sgr A$^*$, the IR upper limits require that
$p$ must be much larger if assuming a power-law form, or, the
distribution is approximately a relativistic Maxwellian form
(YMF02). The discrepancy between NGC~4258 and Sgr A$^*$ may be due to
the different physical parameters at the shock region, especially the
shock strength. On the other hand, due to the poor constraints on the
IR spectrum in Sgr A$^*$, we cannot exclude that some fraction of
electrons may be accelerated into a very steep power-law
distribution.}.  As the result of radiative losses, the {\em steady
distribution} becomes steeper, i.e., $p \rightarrow p+1=3.8$
(Kardashev, Kuz'min, \& Syrovatskii 1962).

The value of $\gamma_{\rm min}$ is important for our spectral
fitting. From Eq. (1), we can only determine the equivalent
temperature of the post-shock plasma, but we do not know the
respective temperature of electrons (or $\gamma_{\rm min}$) and
protons. Another important quantity for determining the radiative
cooling is the strength of the magnetic field. Both of these are
difficult to estimate from first principles. We therefore follow the
usual approach, using two parameters, $\epsilon_e$ and $\epsilon_B$,
to describe respectively the magnitude of the energy densities of
random relativistic electron motions and of the magnetic field in
terms of their ratios to the total thermal energy density of the
post-shock plasma, \be \epsilon_e=\frac{p-1}{p-2}\frac{\gamma_{\rm
min}n_e m_ec^2}{e}, \ee \be \epsilon_B=\frac{B^2}{8\pi e}.  \ee

To calculate the synchrotron and self-Comptonized emission from the
post-shock plasma, we still need to know the profile of the energy
density after the shock front. 
We find to fit the spectrum, the shock must be radiative.  This means
$\epsilon_e \sim 1$ and so the radiative cooling time of electrons due
to synchrotron emission, \be t_{\rm rad} = \frac{3}{4}\frac{8\pi m_e
c}{\sigma_T\gamma_e\beta_e^2B^2} \approx
8\left(\frac{\gamma_e}{100}\right)^{-1}
\left(\frac{B}{1000G}\right)^{-2}{\rm s}, \ee is much shorter than the
dynamical time of the radiative region, \be t_{\rm dyn} \approx
\frac{r_0}{v_2} \approx 10^4\left(\frac{v_2}{0.1c}\right)^{-1}{\rm s},
\ee for typical values of $\gamma_e\sim\gamma_{\rm min}$, $B$ and
$v_2$ in our model.  We find that this holds true even when the
self-absorption of synchrotron emission is taken into account.

In contrast to NGC~4258, the shock in Sgr A$^*$ is adiabatic.  Besides
the difference in mass between the two black holes, which will affect
the dynamical timescale, this is due to the larger magnetic field in
NGC~4258 compared to Sgr A$^*$.  This seems consistent with the idea
of jets having a magnetic origin, because a stronger magnetic field in 
NGC~4258 is consistent with its jet being much stronger.
 
The structure of radiative shocks has been discussed in many papers
(e.g., Blandford \& McKee 1976; Draine \& McKee 1993; Granot \&
K\"onigl 2001).  In addition to a ``shock transition zone'' where the
bulk of the kinetic energy of the pre-shock gas is dissipated and
which can be treated as an infinitesimal front, there exists a thin
``radiative zone'' or ``cooling layer'' after the shock front, where
most of the dissipated energy of electrons is radiated away. In
particular, the magnetic field can be amplified due to the cooling
compression of the gas.  Thus if the pre-shock magnetic field in the
plasma is already sufficiently strong, as in our present case of ADAF,
the magnetic field in the cooling layer can reach or even exceed
equipartition with the electron energy density, i.e., $\epsilon_B \ga
\epsilon_e$ (e.g., Granot \& K\"onigl 2001)\footnote{ Such an
approximate equipartition is also required in the modeling of blazars
(Readhead 1994; Bower \& Backer 1998), microquasars (Kaiser, Sunyaev,
\& Spruit 2000) and GRBs (Granot, Piran, \& Sari 1999).}.

Such a cooling layer is typically removed from the shock front by a
short distance.  An exact approach would be to calculate the radiation
hydrodynamics of the post-shock plasma to obtain the
distance-dependent profiles of $n_e$, $T_e$, and $B$ after the shock
front. We here simply treat the cooling layer as a very thin
homogeneous slab immediately after the shock front.  The emission from
the jet downstream of this slab can be neglected since almost all the
energy of electrons will be radiated away in this thin layer until the
electrons are accelerated again (see Sect. 3.3 below).  The width of
the cooling layer is determined by the post-shock velocity and cooling
time of electrons, \be l_c \approx v_2 t_{\rm rad} = v_2 S l_c
\epsilon_e e/L_{\rm tot} \ll r_0, \ee where $S$ is the ``cross
section'' of the cooling layer.  The effect of the vertical component
of velocity of post-shock plasma is neglected since it should be small
at the beginning of the jet. The shock may be oblique and the shock
normal direction may be between the radial and vertical
directions, and the cross section of the shock front is likely a ring
rather than a full circle (thus the jet may have a cocoon structure).
We here approximate the cross section of the shock as $S=\pi
r_0^2$.  $L_{\rm tot}$ is the emitted total luminosity from electrons,
and is obviously a function of $l_c$, which we therefore calculate
iteratively. Note that we cannot simply use the ratio between the energy
and synchrotron power to calculate the electrons'
radiative lifetime $t_{\rm rad}$ because the effect of synchrotron
self-absorption is significant.  It is not, however, enough of an
effect to change the ``radiative'' feature of the shock.

For a given set of parameters $n, \epsilon_e$ and $\epsilon_B$, we
numerically calculate the emitted luminosity and spectrum 
instead of using the simplified analytical formula 
because of the cutoffs in the emitting particle distributions.
This is because the
minimum frequencies of synchrotron and inverse Compton emission,
$\nu_{\rm syn}(\gamma_{\rm min}) = \gamma_{\rm min}^2 q_e B/2\pi m_ec$
and $\gamma_{\rm min}^2 \nu_{\rm syn}(\gamma_{\rm min})$, fall in the 
mid-IR and soft X-ray bands, respectively, and the slope of the 
inverse Compton emission will thus be changing in the fitted region,
as discussed in the next section.

We would like to note that the number of free parameters in our model
is almost the same as the SSD-ADAF model of GNB99.
There are five free parameters
in our model, namely the shock radius
$r_0$, the mass loss rate in the jet $\dot{M}_{\rm jet}$, the electron energy
spectral index $p$, and the electron/magnetic
energy parameters $\epsilon_e$/$\epsilon_B$. As a comparison,
five free parameters are also required in GNB99, namely
the accretion rate $\dot{M}$, the SSD-ADAF transition radius $r_{\rm tr}$,
the viscous parameter $\alpha$, the magnetic parameter $\beta$, and
the parameter describing the fraction of viscous energy that goes into the
electrons, $\delta$. 

\subsection{Infrared to X-ray Spectral Fitting Results} 

\begin{figure}
\psfig{file=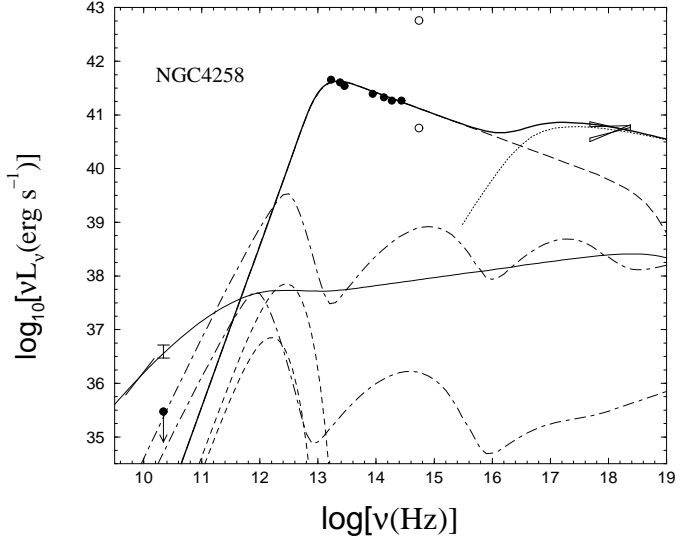,width=8.8cm,angle=270}
\caption{Model fit to the spectrum of NGC~4258. Parameters are chosen
to fit the {\em Beppo}SAX data, which is described in Sect. 2. The
thick solid line shows the emission from the radiative shock at the
base of the jet, with the long-dashed and dotted lines being for the
synchrotron and its self-Comptonized emission, respectively (Sect. 3.1
\& 3.2).  The thin solid line shows the emission from the outer part
of the jet (Sect. 3.3). The dot-dashed and short-dashed lines are for
the underlying ADAF and SSD, respectively, with two different
accretion rates, and are minor contributors to the spectrum.}
\end{figure}

As we stated in Sect. 2, the X-ray spectral slope varies between the
different observations, e.g., $s=0.64 \pm 0.08$ (Pietsch \& Read
2002), $0.78\pm 0.29$ (Makishima et al. 1994), $0.87\pm0.15$ (Reynolds
et al. 2000), and $1.11 \pm 0.14$ (Fiore et al. 2001).  In addition to
instrumental reasons, it is very possible that such variations are
intrinsic, as we know happens in some sources (e.g., Giebels et
al. 2002).  We will therefore fit both the steepest ($s=1.11$, from
{\em Beppo}SAX) and the flattest ($s=0.64$, from XMM-{\em Newton})
X-ray data in the framework of our model.

Fig. 1 shows our best fit to the {\em Beppo}SAX results. The
parameters are $\dot{M}_{\rm jet}=2.9\times 10^{-4}\msunyr$,
$r_0=5r_g$, $\epsilon_e=0.27$ and $\epsilon_B=0.31$. From these
parameters we obtain $B=1400$G, $\gamma_{\rm min}=30$, the length of
the cooling layer $l_c=0.01r_g$, and the particle density $
n=1.8\times 10^9~{\rm cm}^{-3}$.  The long-dashed line in the figure
shows the synchrotron emission, the dotted line shows its
self-Comptonized radiation, and the thick solid line shows their sum.
The power-law IR spectrum with $s=1.4$ is fitted by the synchrotron
emission from electrons with a steady energy distribution of index
$p=3.8=2s+1$. The predicted flux at the optical band is also roughly
in agreement with the lower flux estimated in W95, which we feel is
more likely (see Sect. 2.3).  In the radio band, due to the strong
synchrotron self-absorption and the fact that the emission from the
region out of the cooling layer can be neglected, the synchrotron
spectrum steeply decreases below the synchrotron peak so the
$220~\mu$Jy upper limit at 22~GHz is easily satisfied.

The X-ray spectrum is mainly due to the synchrotron self-Compton
component.  Due to the effect of the low-energy cutoff in the electron
distribution, the power-law spectrum at $2-10$~keV is harder than
$s=(p-1)/2=1.4$. The exact value of the spectral index depends on the
minimum frequency of inverse Compton emission, $\sim \nu_{\rm
syn}^p\gamma_{\rm min}^2$, here $\nu_{\rm syn}^p$ is the peak
frequency of the synchrotron emission,
which is mainly determined by the value of $B$ at the cooling
layer. If $\nu_{\rm syn}^p\gamma_{\rm min}^2$ is smaller, the spectrum
would become steeper.

Also shown in Fig. 1 is the emitted spectrum from the underlying
accretion disk, namely the ADAF (dot-dashed lines) and SSD
(short-dashed lines).  Because the value of the accretion rate is
uncertain, we here adopt two values: $\dot{M}=\dot{M}_{\rm NM95}$ (for
the lower set of lines), and $10\dot{M}_{\rm NM95}$ (for the upper set
of lines).  The transition radius is set to be $r_{\rm tr}=$0.13pc,
but our results are not sensitive to its exact value. The spectrum of
ADAF is calculated by self-consistently solving the radiation
hydrodynamic equations (e.g. Yuan et al. 2000).  The ADAF with
$\dot{M}=\dot{M}_{\rm NM95}$ produces very low radio flux at 22~GHz
because of small accretion rate, well satisfying the 220~$\mu$Jy upper
limit.  The accretion rate $10\dot{M}_{\rm NM95}$ is the largest rate
for which an ADAF can satisfy the 22~GHz upper limit for $r_{\rm tr}=
0.13$~pc
\footnote{The accretion rate can be even higher if the transition
radius is smaller, as in GNB99.}. In both cases, the contributions of
ADAF and SSD to the spectrum of interest are rather small and can even
be neglected. Decreasing $r_{\rm tr}$ will increase the contribution
of SSD, but we find even for $r_{\rm tr}\approx 50r_g$---which is
prohibited by the iron line observation---the contribution of SSD is
still too small to contribute to the spectrum for $\dot{M}\la
10\dot{M}_{\rm NM95}$.

\begin{figure}
\psfig{file=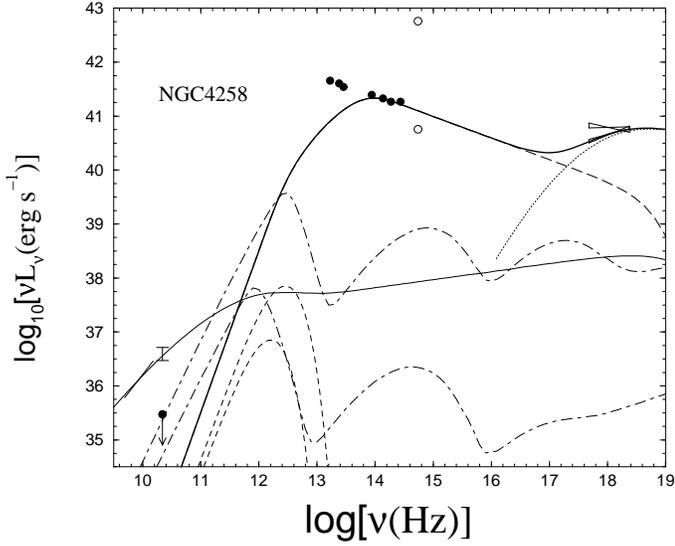,width=8.8cm,angle=270}
\caption{Model fit to the spectrum of NGC~4258. The parameters are
chosen to fit the XMM-{\em Newton} result.  The lines are defined as
in Fig. 1. The discrepancy in the mid-IR may be due to variability
(see text for details).}
\end{figure}

Fig. 2 shows our results of fitting the XMM-{\em Newton} X-ray
data. The parameters are $r_0=3r_g$, $\dot{M}_{\rm jet}=7\times
10^{-5}\msunyr$, $\epsilon_e=0.2$, and $\epsilon_B=0.11$. Such
parameters give $B=1250$G, $\gamma_{\rm min}=75$, $n=1.2\times
10^9~{\rm cm}^{-3}$, and $l_c=0.004r_g$. The harder inverse Compton
spectrum compared to Fig. 1 is due to the larger $\nu_{\rm
syn}^p\gamma_{\rm min}^2$.  Thus, we interpret the variations in the
X-ray spectrum of NGC~4258 as due mainly to variations of $\gamma_{\rm
min}$, which could result physically from changes in the shock
location $r_0$.  However, compared to Fig. 1, this set of parameters
significantly underpredicts the mid-IR flux of Chary et al. (2000).
This may be due to intrinsic variability in NGC~4258, since the
broadband data is not simultaneous.  According to our model, when the
X-ray spectrum becomes harder, the mid-IR flux is decreased due to the
increased self-absorption from the required lower-energy electrons.
Our results would therefore be consistent if the high mid-IR flux
detected by Chary et al. (2000) was obtained when the X-ray spectrum
was as steep as the {\em Beppo}SAX result.  Future simultaneous
IR and X-ray observations would be a good test of this model.

The X-ray emission in our model comes from small spatial scales,
$r_0\sim 3-5r_g$, and therefore rapid variability as detected in Fiore
et al . (2001) would be expected. On the other hand, we can easily set
$r_{\rm tr}>100r_g$ so our model will not produce a broad iron line,
consistent with the observation of Reynolds, Nowak, \& Maloney (2000).

In our model presented in Fig. 1., $\dot{M}_{\rm jet}\approx 4
\dot{M}_{\rm NM95}$.  If the accretion rate in the inner ADAF really
equals $\dot{M}_{\rm NM95}$, this result is unphysical. It is possible
that this factor of $\sim 4$ discrepancy could be absorbed by the
uncertainties in both the NM95 model and our own\footnote{For example,
in addition to other simplifying approximations in our model, we
assume that the proton density is the same with the electron density,
$n_p=n_e$ when calculating $n_e$ from $\dot{M}_{\rm jet}$.  It is also
possible that proton-proton collisions in the post-shock plasma, and
the ensuing pion creation and decays, would lead to significant
amounts of additional $e^{\pm}$.  In this case $n_p < n_e$, and the
required $\dot{M}_{\rm jet}$ would become smaller.  We are
investigating this possibility elsewhere (Markoff et al. in
preparation).}. However, as we show above, we can still be under the 22
GHz constraint with an accretion rate as high as $10\dot{M}_{\rm
NM95}$.  In this case, a fraction of $25\%$ (for Fig. 1) and $10\%$
(for Fig. 2) of accretion flow will be transferred into the jet. We
speculate that the latter is more likely.

We now discuss why we require the accretion flow inside $r_{\rm tr}$
to be an ADAF (or more generally a radiatively inefficient accretion
flow) rather than a SSD. To explain the spectrum of NGC~4258,
especially the power-law IR spectrum whose luminosity is almost equal
to the bolometric luminosity of NGC~4258, extremely relativistic
electrons with power-law energy distribution are required.  An ADAF is
basically adiabatic, i.e., the potential energy of the accretion flow
is stored in the protons without being radiated away.  When the flow
is shocked at $r_0$, the stored energy in the pre-shock ADAF will be
transferred into the thermal energy of post-shock plasma. Among them a
large fraction, $\epsilon_e\approx 0.2-0.3$, will be used to
accelerate electrons into a relativistic power-law distribution. Note
that the radiative efficiency of the radiative shock is much higher
than an ADAF, which follows, \be \eta \equiv
\frac{L_{bol}}{\dot{M}_{\rm jet}c^2} \sim \frac{\epsilon_e
GM\dot{M}_{\rm jet}}{(r_0-2r_g)\dot{M}_{\rm jet}c^2}=
\frac{\epsilon_e}{r_0/r_g-2}\approx 0.1-0.2.  \ee Here we simply use
the Paczy\'nski \& Wiita (1980) potential to mimic the geometry of the
central black hole.  If the ADAF were replaced by a SSD, the accretion
energy would be radiated through a large range of radii.  Therefore,
even though a shock would still occur, only a small fraction of
accretion energy would go into the thermal energy of the post-shock
plasma. The electrons would not obtain a relativistic power-law energy
distribution and their radiative efficiency would be $\ll 0.1$.

There exists another possible piece of evidence for the existence of
an ADAF rather than a SSD within $0.13~$pc.  Observation indicated
that within this radius the water maser emission disappears. In order
to account for the absence of masers, NM95 propose that X-ray
irradiation ``shuts off'' inside this radius.  If this is true, Menou
\& Quataert (2001) show that for an accretion rate 
as low as $\dot{M}_{\rm NM95}$,
both MHD turbulence and gravitational instability will cease to be a
viable angular momentum transport mechanism, therefore, the accretion
in NGC~4258 should proceed very inefficiently or perhaps not at
all. This excludes the possibilities of producing the bolometric
luminosity of NGC~4258 through a SSD. The
absence of masers may be due to the transition between SSD and ADAF.
Thus the X-rays will irradiate both the SSD and the ADAF, and MHD
turbulence will still be an efficient angular momentum transport
mechanism even though the accretion rate in NGC~4258 is so low (Menou
\& Quataert 2001).

\subsection{Radiation from the Outer Part of the Jet and Radio Observations}

According to the above scenario, a large fraction of accretion flow
will be ejected out of the ADAF and go into the jet after passing
through a radiative shock. The emission from the thin cooling layer
downstream of the shock front explains the IR and X-ray spectra of
NGC~4258.  We now follow the propagation of the plasma and discuss the
emission from the jet downstream of the cooling layer. We consider
only the northern jet in NGC~4258 since the southern jet is severely
affected by the maser disk. If our model is correct, the propagated
jet should be able to fit the radio spectrum detected by Turner \& Ho
(1994), Herrnstein et al. (1998), and the 22 GHz offset. We explain
this by the canonical jet model of Blandford \& K\"onigl (1979)
(see also Falcke \& Biermann 1999). The numerical
approach is presented in Falcke (1996) and Falcke \& Markoff (2000).
The plasma will
become supersonic after being accelerated in a ``nozzle''.  The
velocity profile along the jet is then calculated by solving the
relativistic Euler equation. The opening angle of the jet is
determined by the Mach number of the jet, $\varphi \sim M^{-1}$. Given
the mass loss rate in the jet determined at the base of the jet,
$\dot{M}_{\rm jet}$, the density profile is then determined.  After
the radiative zone, the jet plasma initially emits very little since
the electrons are very cool after radiating most of their energy away.
However, we assume that after propagating a certain distance from the
nucleus, $z_0$, the electrons in the jet will be accelerated through,
e.g., the internal shocks due to the velocity irregularities in the
jet beam (Rees 1978).
The electrons will then be accelerated again into a power-law form,
$N_{e,j}(\gamma_{e,j})=N_{e,j}^0\gamma_{e,j}^{-p}$, with $\gamma_{e,j}
\ge \gamma_{\rm j,min}$ and taking the standard strong shock value
$p=2.5$, although the model is not sensitive to the exact value.  The
average energy of electrons (or equivalently $\gamma_{\rm j,min}$) can
be calculated by the relativistic shock transition condition, \be
\frac{p-1}{p-2}\gamma_{j,{\rm min}}m_ec^2=\epsilon_{j,e}
(\Gamma_j(z_0)-1)m_pc^2.  \ee Here $\Gamma_j(z_0)$ is the bulk Lorentz
factor of the jet at $z_0$ and $\epsilon_{j,e}$ is the energy density
factor of electrons.
 
The shock location $z_0$ must be $\gg r_g$. 
Herrnstein et al. (1998) obtained the 22~GHz upper limit by extracting
the emission located within an $\sim 0.24\times 0.24$~mas box centered
on the peak of the northern jet emission located at $\sim 0.4~$mas.
This requires that the contribution to the 22~GHz flux from the jet
within the region $0< z < (0.4-0.24/2)~ {\rm mas} \simeq 5600r_g$ must
be below the 220$\mu$Jy upper limit. We therefore set $z_0=6000r_g$,
but again the fit is not sensitive to its precise value.

Our calculation indicates that $\Gamma_j(6000r_g)\approx 3$ and
$\gamma_{j,{\rm min}}\approx 650$ for $\epsilon_{j,e}=0.5$.  Further
assuming equipartition between the electrons and magnetic energy, and
a viewing angle of $\theta=83^\circ$ (Miyoshi et al. 1995), we can
calculate the emergent spectrum of the jet and the location of the
peak, $z(\nu)$, of the emission at a frequency $\nu$.  The result is
shown by the thin solid line in Fig. 1.  We see that it fits the
results of Turner \& Ho (1994) and Herrnstein et al. (1998) very well
\footnote{We find the synchrotron emission in the jet in NGC~4258 is
optically thin. This can be understood directly from observations
(GNB99). Observation indicated that the opening angle of the jet is
about $10^{\circ}-15^{\circ}$ (Herrnstein et al. 1997). The radio
brightness temperature is $\sim 10^9$K, which is much smaller than the
kinetic temperature of electrons responsible for the emission, $T_e
\sim \gamma_{j,{\rm min}}m_ec^2/3k\approx 10^{12}$K.}.  Moreover, in
our result, the 22 GHz emission mainly comes from the location $\sim
0.4~$mas from the nucleus, satisfying the observed (time variable)
$0.35-0.46$~mas offset (Herrnstein et al. 1998). We explain the
variation of the offset with time as due to the variation of the shock
location $z_0$.  If $z_0 \ll 6000r_g$, to fit both the 22GHz flux and
its offset simultaneously, a much smaller $\dot{M}_{\rm jet}$ would be
required. But in this case, the synchrotron emission from the jet
within the region $0<z<z_0$ would greatly exceed the 0.22 mJy upper
limit. This supports our adopted $\dot{M}_{\rm jet}$, and further, our
jet-ADAF model of NGC~4258.

As shown by the thin solid line in Fig. 1, the contribution from the
outer part of the jet to the X-ray flux can be neglected compared to
the cooling layer at the base of the jet.  NGC~4258 is the only source
for which, to our knowledge, both the offset and an upper limit of the
22~GHz emission from the nucleus are so precisely determined. This
suggests that it may be possible that the X-ray spectra of some
sources which were previously attributed to emission from the outer
part of the jet actually comes from the base of the jet (or in other
words, the coupling region between the jet and accretion disk).  Of
course, an additional important factor when we compare the respective
contribution of the outer jet and base of jet is the Doppler factor
$\delta$ ($\delta=\left[\Gamma_j(1-\beta_j{\rm
cos}\theta)\right]^{-1}$).  At the base of the jet, $\delta \approx
1$, but for the outer part of the jet, $\delta$ in general deviates
from 1 significantly, depending on the value of $\theta$.  For
NGC~4258, $\theta\approx 83^{\circ}$, so $\delta \ll 1$, and the
emission from the base of the jet dominates over that from the outer
part of the jet. But for some sources like Blazars,
$\delta \gg 1$, thus the emission from the outer part of the jet may
dominate over that from the base of the jet.  Taking the effect of
$\delta$ into account, the candidates in which emission from the base
of the jet may dominate over the outer jet are those having misaligned
jets.  M87 will be a good example.
 
\subsection{Model predictions}

Taking into account the emission from the outer part of the jet and
the base of the jet, our model makes the following predictions for NGC~4258:
\begin{itemize}
\item
There exists an infrared bump peaked at
$\sim 50 \mu$m. The radio spectrum should be flat with spectral index of
$s\approx 0.3$ until a break frequency $\nu_b\approx 200-400{\rm GHz}$,
with the exact value depending on the accretion rate of the underlying ADAF. 
These results can be tested by the future ALMA telescope.
\item
Rapid variability at IR band with timescale of $\sim 1$~hr is 
expected as detected in X-ray band, and the IR and X-ray variability 
should be tightly correlated.
\item
A somewhat weaker prediction of our model is 
that, if the mass ejection from the accretion disk suffers the same 
kind of ejection-related instability as in microquasar GRS 1915+105 ( 
Dhawan, Mirabel, \& Rodriguez 2000) and maybe also 
Sgr A$^*$ (Zhao et al. 2001),
a similar quasi-periodic variability at submm-IR bands should be detected.
The period in the case of Sgr A$^*$ is about four months 
(Zhao et al. 2001; Yuan \& Zhao 2002). Scaled with the mass
of the black holes, we predict a variability period of about $4\sim5$ yrs
at submm-IR bands in NGC~4258 .
\item
Extending these predictions to other sources, we will argue in the
next section that the spectra of some LLAGNs may have the same origin
as NGC~4258. In this case we would predict strong correlations between
the radio, submm, IR, and X-ray emission in these sources.
\end{itemize}

\section{Discussion: is NGC~4258 a typical LLAGN?}

Recently much attention has been paid to LLAGNs, a class which
includes many of our nearby galactic neighbors.  This class of AGN is
very common in the local universe, e.g., 43\% of all northern galaxies
brighter than $B_T=12.5$ mag are active in the form of emission-line
nuclei classified as Seyferts or LINERs (Ho, Filippenko \& Sargent
1997; Ho et al. 2001). The spectral energy distribution of LLAGNs are
markedly different with the canonical luminous AGNs which are usually
associated with the big blue bump (e.g., Elvis et al. 1994). Ho (1999)
presented the spectra of a sample of seven LLAGNs for which reasonably
secure black hole masses have been determined by dynamical
measurements.  Several most obvious features of their spectra are as
follows (Ho 1999).

\begin{itemize}
\item 
The optical/UV slope is quite steep. The averaged power-law index is 
1.5, if two possibly highly reddened objects are excluded, while in 
canonical AGNs it is 0.5-1.0.

\item 
The UV band is exceptionally dim relative to the optical and X-ray bands.
There is no evidence for a big blue bump component in any of the objects.
The spectrum reaches a local minimum somewhere in the far-UV or extreme-UV
region. 

\item If fitting the X-ray spectra with a power-law
($f_{\nu}\propto\nu^{-s}$),
the spectral index $s$ obtained is very variable among the sources, 
$s \approx 0.6-1.2$
\footnote{After the publication of Ho (1999) paper, new X-ray observational
data of two sources are available, namely $s=1.2$ for M87
(B\"ohringer et al. 2001; Wilson \& Yang 2002), $s=0.88$ for NGC4579
(Eracleous et al. 2002), and $s=0.94$ for M81 (Page et al. 2002).}.

\item There is tentative evidence for a maximum in the spectrum at 
mid-IR or longer wavelengths. 

\item The nuclei have radio spectra that are either flat or inverted. 

\end{itemize}

Ho (1999) argued that the absence of the big blue bump is a property
intrinsic to the nuclei and not an artifact of strong dust
extinction. In addition to the above spectral features, the iron line
in LLAGNs is in general extremely weak or absent (Terashima et
al. 2002).

In contrast with the obvious importance of LLAGNs, we are still not
clear about their physics. From their low luminosity in units of
Eddington luminosity, the absence of big blue bump, the weak or absent
iron line, and other observational facts, we can deduce that the inner
region of the accretion disk may be best described as an ADAF rather
than SSD (e.g., Ho 2002). However, it is very hard for an ADAF to fit
the spectra of LLAGNs.  Quataert et al. (1999) use an SSD-ADAF
transition model to fit the spectra of two LLAGNs, namely M81 and
NGC4579.  In addition to underpredicting the radio spectrum, which
they attribute to the ``contamination by jets'', the predicted
optical/UV spectra are somewhat too steep. This is because in their model the
optical/UV is mainly produced by a SSD whose spectrum is exponentially
decreasing at this waveband.  It may be even more challenging to fit
the spectra of other LLAGNs in the sample of Ho (1999) with such a
model.

On the other hand, the excellent agreement between the spectra of
NGC~4258 and of general LLAGNs indicates that NGC~4258 is likely a
typical LLAGN.  In fact, it seems to be the LLAGN for which we have
the best data so far, although Sgr A$^*$ is a close second. NGC~4258
is better, however, since in addition to the precise determination of
the mass of the black hole, the distance to the source, and the
broadband spectrum ranging from radio to hard X-ray, NGC~4258 has
additional extremely important constraints.  Specifically, the 220
$\mu$Jy upper limit and the offset at 22GHz may provide ways to
determine the mass accretion rate from the modeling of the water maser
emission. Thus the success of our jet-ADAF model in explaining the
spectra of both NGC~4258 and Sgr A$^*$ may be relevant to our
understanding of other LLAGNs.

We therefore suggest that the spectra of LLAGNs are at least partly
due to the emission of the jet, namely that the synchrotron emission
from the base of the jet is responsible for the steep IR/optical/UV
spectrum peaked roughly at mid-IR, and its self-Comptonized emission
produces the X-ray spectrum.  The flat/inverted radio spectrum comes
from further out in the jet.  In the radio wavebands, this seems to be
already well-established (Falcke 2001; Ulvestad \& Ho 2001; Nagar et
al 2001). Several important results of radio observations, i.e., high
brightness temperature, the flat spectra, and the compactness of the
radio cores, all are best explained by the emission from the jet.  In
the X-ray wavebands, the variety of the X-ray power law spectral
indexes can be explained as due to the variety of both the peak
frequency of synchrotron emission-$\nu_{\rm syn}^p$-and $\gamma_{\rm
min}$ among the sources, similar to the variations in the X-ray
spectrum of NGC~4258 as we discussed in Sect. 3.2.  The variety of
optical/UV power-law indexes may be the reflection of the complexity
of shock type (radiative or adiabatic) and shock acceleration (the
value of $p$).

The uncertainty in the extent of the underlying disk contribution adds
complexity to this picture.  It is possible that, at some wavebands or
for some sources the emission from the underlying ADAF, and maybe also
SSD, will be comparable with or even dominate over that from the jet.
For example, observations indicate that X-ray variability on
timescales less than a day is lacking in many LLAGNs although
short-term variability has been detected in some LLAGNs (Ptak et
al. 1998; Terashima et al. 2002).  This may be an evidence for the
dominance of ADAFs over jets at X-ray band in these sources because in
an ADAF the X-ray emission can be dominated by bremsstrahlung process
whose corresponding variability timescale is very long. The relative
contribution of ADAF and SSD depends on the transition radius between
them, while the relative contribution of the jet and accretion disk
depends on our viewing angle with respect to the jet axis and the
fraction of the accretion flow transferred into the jet.

\begin{acknowledgements}
F. Y. thanks R. Chary for discussions on the infrared observation
results of NGC~4258. The comments of Bozena Czerny, Charles Gammie,
Yuichi Terashima, and Andrew Wilson are also acknowledged.
This work is partially supported by China 973
Project under NKBRSF G19990754 (F.Y.) and AUGER Theory Grant
O5CU1ERA/3 from the BMBF (S.M.). 
\end{acknowledgements}

\end{document}